\documentclass[journal,twoside]{IEEEtran}

\usepackage{cite}
\usepackage{amsmath,amssymb,amsfonts}
\usepackage{graphicx}
\usepackage{textcomp}
\usepackage{multirow}
\usepackage{booktabs}
\usepackage{array}
\usepackage{tabularx}
\usepackage{algorithm}
\usepackage{algpseudocode}
\def\BibTeX{{\rm B\kern-.05em{\sc i\kern-.025em b}\kern-.08em
    T\kern-.1667em\lower.7ex\hbox{E}\kern-.125emX}}
    
\begin{document}

\title{Macroscopic EEG Reveals Discriminative Low-Frequency Oscillations in Plan-to-Grasp Visuomotor Tasks}

\author{
Anna Cetera, \and
Sima Ghafoori, \and
Ali Rabiee, \and
Mohammad Hassan Farhadi, \and
Yalda Shahriari, \and
 Reza Abiri
\thanks{This research study was supported by the NSF CAREER under award ID 2441496 and the NSF grant under award ID 2245558. Additionally, the project was supported by URI Foundation Grant on Medical Research and the Rhode Island INBRE program from the National Institute of General Medical Sciences of the NIH under grant number P20GM103430. All authors are with the Department of Electrical, Computer, and Biomedical Engineering, University of Rhode Island, Kingston, USA. 
(annacetera@uri.edu; sima.ghafoori@uri.edu; ali.rabiee@uri.edu; mh.farhadi@uri.edu; yalda\_shahriari@uri.edu; reza\_abiri@uri.edu).}
% \thanks{Manuscript received XXX XX, 2025; revised XXX XX, 2025.}
}

\maketitle

\begin{abstract}
\textit{Objective:} The vision-based grasping brain network integrates visual perception with cognitive and motor processes for visuomotor tasks. While invasive recordings have successfully decoded localized neural activity related to grasp type planning and execution, macroscopic neural activation patterns captured by noninvasive electroencephalography (EEG) remain far less understood. \textit{Methods:} We introduce a novel vision-based grasping platform to investigate grasp-type-specific (precision, power, no-grasp) neural activity across large-scale brain networks using EEG neuroimaging. The platform isolates grasp-specific planning from its associated execution phases in naturalistic visuomotor tasks, where the Filter-Bank Common Spatial Pattern (FBCSP) technique was designed to extract discriminative frequency-specific features within each phase. Support vector machine (SVM) classification discriminated binary (precision vs. power, grasp vs. no-grasp) and multiclass (precision vs. power vs. no-grasp) scenarios for each phase, and were compared against traditional Movement-Related Cortical Potential (MRCP) methods. \textit{Results:} Low-frequency oscillations (0.5-8 Hz) carry grasp-related information established during planning and maintained throughout execution, with consistent classification performance across both phases (75.3-77.8\%) for precision vs. power discrimination, compared to 61.1\% using MRCP. Higher-frequency activity (12-40 Hz) showed phase-dependent results with 93.3\% accuracy for grasp vs. no-grasp classification but 61.2\% for precision vs. power discrimination. Feature importance using SVM coefficients identified discriminative features within frontoparietal networks during planning and motor networks during execution. \textit{Conclusion:} This work demonstrated the role of low-frequency oscillations in decoding grasp type during planning using noninvasive EEG. \textit{Significance:} These findings provide a foundation toward scalable, intention-driven Brain-Machine-Interface (BMI) control strategies.
\end{abstract}


\section{Introduction}
\label{sec:introduction}

\IEEEPARstart{R}{each-to-grasp} disabilities in individuals with severe paralysis, such as spinal cord injury (SCI), have a profound impact on their quality of life. With approximately 370,000 SCI patients in the United States and 18,000 new cases reported annually, restoring arm and hand functionality is a top priority to improve independence in activities of daily living (ADLs)\cite{wolpaw2004_control, pistohl2012_decoding}. For individuals with SCI, spinal cord lesions interrupt the descending corticospinal pathways that propagate grasp commands from the brain to the spinal motor neurons, preventing the execution of associated reach-to-grasp tasks. However, cortical and subcortical motor planning networks above the lesion remain intact, preserving the ability to formulate a grasp command prior to grasp execution\cite{chen2016_functional}.

Invasive electrophysiological studies in non-human primates showed that grasp planning occurs several hundred milliseconds before execution~\cite{schaffelhofer2016_object,vargas-irwin2015_linking} with localized anterior intraparietal area (AIP) and F5 containing canonical neurons that encode specific grasp types even during passive object viewing~\cite{murata2000selectivity,rizzolatti2014cortical}. Similarly in humans, the vision-based grasping network integrates ventral stream object recognition with dorsal stream visuomotor transformation, with the localized anterior intraparietal sulcus (aIPS) and ventral premotor cortex (PMv) showing grasp-selective activity during object observation and motor planning~\cite{chinellato2009_neuroscience}. 

Within these brain regions, studies have demonstrated that neural oscillations show frequency-specific functional roles during visuomotor tasks. Low frequency oscillations ($< 8$ Hz) have been used to decode grasp type in invasive studies~\cite{natraj2021compartmentalized,hall2014_corticalsleep} and are associated with cognitive control~\cite{cavanagh2014frontal, tan2024theta} and motor decision-making~\cite{murray2025role} recruited for grasp planning. Higher frequencies, mainly emerge during movement execution and are associated with sensorimotor integration~\cite{pfurtscheller2003motor} and attention to task engagement~\cite{dreyer2023_graspspecific, khanna2015_betamotor}. These findings establish that grasp planning and execution may operate through large-scale and distinct neural mechanisms that are temporally and spectrally separable. While invasive methods provide high spatial and temporal resolution within localized brain regions~\cite{shenoy2013_cortical}, they are limited in capturing the macroscopic emergence of plan-to-grasp related neural activity that EEG provides through its complete electrode coverage across all brain regions.

Despite the understanding of grasp planning at the invasive-level, current noninvasive EEG-based grasp decoding studies have predominantly focused on reach-to-grasp execution, leaving the planning phase largely unexplored~\cite{agashe2015_global, jochumsen2016_eegdetecting}. Many existing analysis techniques rely on Movement-Related Cortical Potentials (MRCPs), which capture a window of neural activity across planning and execution phases, preventing the isolation of planning-specific neural dynamics ~\cite{schwarz2020_analyzing, sun2024_unraveling}. From an experimental paradigm perspective, these studies often present multiple objects simultaneously~\cite{xu2021_decoding}, use screen-based observation~\cite{hooks2023_eegdecoding, sburlea2021_disentangling}, or design custom-made objects solely for grasp decoding~\cite{iturrate2018_eeggrasp} rather than naturalistic objects involved in ADL. These paradigms are limited by the absence of an isolated grasp planning phase to investigate whether the associated macroscopic EEG neural activity can discriminate grasp types.

To address these limitations, we developed a novel vision-based grasping platform that temporally isolates planning from execution phases during naturalistic grasp tasks with familiar ADL objects~\cite{ghafoori2024_bispectrum, rabiee2024_wavelet}.  By presenting objects that appear in ADL to participants, this paradigm is designed to recruit preserved cognitive mechanisms that remain functional in individuals with motor impairments~\cite{bienkiewicz2014tool}. Filter-Bank Common Spatial Pattern (FBCSP) was utilized as the primary analysis method due to its success in identifying discriminative frequency-specific features in motor imagery tasks~\cite{ang2012_fbcsp, ray2015_a} and motor execution studies~\cite{chaisaen2020_decoding, mohseni2020_upperwaveletcsp, shiman2017_classification}. Based on these prior findings and given that both motor imagery and grasp planning are cognitive tasks performed without physical movement, we hypothesized that FBCSP can capture the emergence of frequency-specific neural patterns during plan-to-grasp tasks at the macroscopic level through noninvasive EEG. Our specific contributions include:

\begin{enumerate}
   \item Development of a novel EEG-based platform that temporally isolates grasp planning from execution phases during naturalistic object interactions.
   \item Comprehensive macroscopic-level analysis using FBCSP to explore discriminative frequency-specific neural activity associated with grasp planning and execution phases, with established MRCP methods used for performance comparison. 
   \item Distinctive role of low-frequency oscillations in discriminating precision from power grasps during the planning phase, allowing for a more naturalistic control of BMI systems.  
\end{enumerate}

By prioritizing grasp decoding during the planning phase, our platform bridges the gap for real-world neuroprosthetic applications, offering a scalable, noninvasive control strategy to improve functional independence for individuals with severe paralysis.

\section{Materials and Methods}

\subsection{Experimental Setup and Task Design}
\subsubsection{Experimental Setup} 
\begin{figure*}[!t]
\centering
\includegraphics[width=0.9\textwidth]{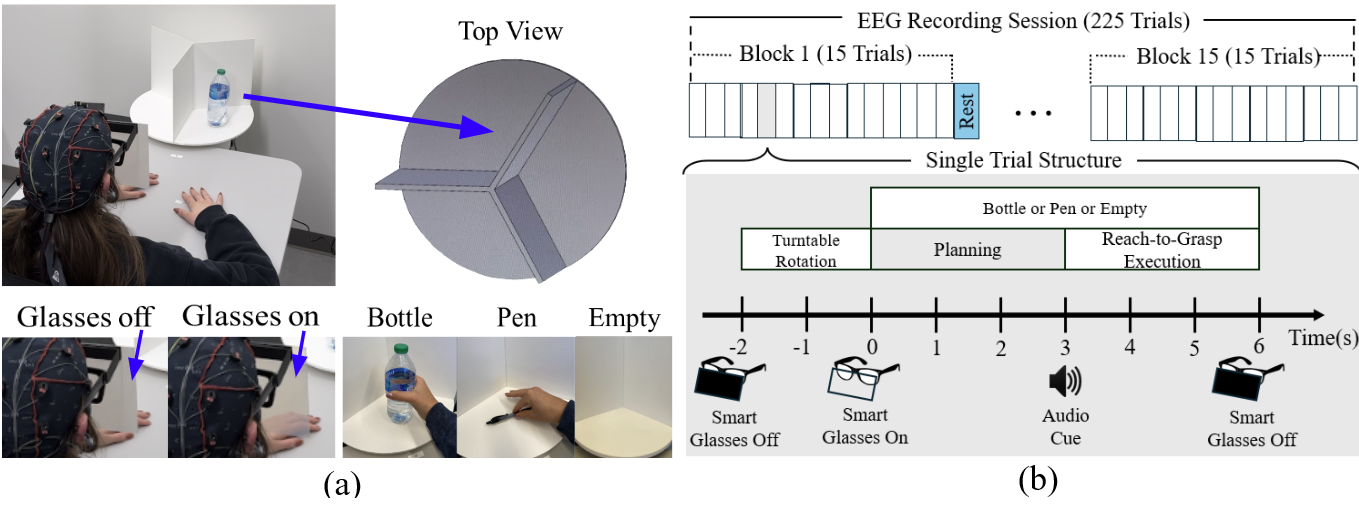}
\vspace{-8pt}
\caption{Experimental setup and data collection protocol for vision-based grasping platform. (a) Depicts a participant wearing the 16-channel EEG recording system during the experimental protocol. The top view of the turntable rotates to present one of three conditions at random: bottle, pen, or empty. (b) Single-trial structure and outline of EEG recording session, beginning with the planning phase, then the reach-to-grasp execution phase following audio cue.}
\label{fig:experimental_setup}
\end{figure*}

The experimental platform consisted of a PC-controlled motorized turntable with three isolated sections (120°), presenting ADL objects (pen for precision grasp, water bottle for power grasp) or an empty control condition to participants positioned 30 cm away (Fig.~\ref{fig:experimental_setup}). Smart glasses with Polymer Dispersed Liquid Crystal (PDLC) film controlled object viewing by transitioning between opaque and transparent states to limit anticipatory bias in upcoming trials, allowing neural activity to capture task-relevant processes~\cite{fehrer1962reaction}. Detailed hardware and specifications are described in Cetera et al.~\cite{cetera2024_emerging}.

\subsubsection{Task Design and Protocol}  
Participants completed 225 plan-to-grasp trials (75 per condition: pen, bottle, empty) using an audio-cue based paradigm (Fig.~\ref{fig:experimental_setup}). Participants sat upright with palms positioned at predefined locations 30-cm from the turntable center. Each 6-second trial began with smart glasses transitioning from opaque to transparent, allowing 3 seconds of object observation (planning phase). A 250-ms auditory beep cued participants to initiate reach-to-grasp movements, providing 2.75 seconds for controlled, natural-paced execution. For empty trials, participants observed the empty section but performed no reach-to-grasp. Following practice sessions to familiarize participants with the experimental timing and procedures, data collection proceeded in blocks of 3-15 randomized runs (3 trials each run) with rest periods between blocks to prevent fatigue.

\vspace{-10pt}
\subsection{Data Acquisition and Preprocessing}
For EEG recording, 16 active channels were used with a biosignal amplifier (g.tec medical engineering, Austria). The electrodes were positioned across all brain regions following the 10-20 international system. The EEG data was recorded at a sampling rate of 256 Hz, followed by an offline zero-phase 4th-order Butterworth bandpass filter with cut-off frequencies at 0.5 to 40 Hz. Eye movements were minimized as the participants' gaze was fixed along the same visual axis towards the presented object during each trial. Eye blink and muscle artifacts were removed using Infomax Independent Component Analysis (ICA) with components rejected based on visual inspection of characteristic artifact patterns for each trial. Lastly, the single trial preprocessed data was sorted according to class labels: pen (class ID: 0), bottle (class ID: 1), and empty (class ID: 2). All bandpass and lowpass filters used for all subsequent FBCSP feature extraction were zero-phase, 4th-order Butterworth filters designed with varying cutoff frequencies using a zero-padding buffer at trial boundaries to eliminate edge artifacts.

\vspace{-10pt}
\subsection{Participants}
All participants (5 male, 7 female) were recruited from the University of Rhode Island between the ages of 21 to 38 years and had no history of motor deficits or neurological impairments. Before participation, they were informed of the experimental protocol and provided written informed consent. Participants received monetary compensation for their participation. This study was approved by the University of Rhode Island Institutional Review Board (IRB \#1944644). 

\vspace{-10pt}
\subsection{Filter Bank Common Spatial Pattern (FBCSP) Feature Extraction}

\subsubsection{Dataset Preparation and Splitting}
The preprocessed EEG dataset was split 80/20 for training and testing. Test data comprised 15 randomly selected trials per class, held out for final evaluation:
\begin{itemize}
    \item Pen: 60 training/validation + 15 test = 75 total trials
    \item Bottle: 60 training/validation + 15 test = 75 total trials
    \item Empty: 60 training/validation + 15 test = 75 total trials
\end{itemize}
The remaining 180 trials were used for CSP filter derivation and SVM classifier training/validation.

\subsubsection{Frequency Band Decomposition and Phase Segmentation}
The FBCSP approach was used to analyze EEG data and extract features for classification across multiple frequency bands. The class-sorted EEG data (pen, bottle, empty) were decomposed into the following five frequency bands: delta (0.5-4 Hz), theta (4-8 Hz), alpha (8-13 Hz), beta (13-30 Hz), and gamma (30-40 Hz). Single-trial EEG within each frequency band was segmented into planning and reach-to-grasp phases based on time-stamped event labels collected during the experiment. The filtered signals were processed using the Common Spatial Pattern (CSP) algorithm to extract frequency-specific spatial features for classifying between specific binary and multiclass scenarios.

\subsubsection{FBCSP Formulation}
The CSP algorithm was applied independently to each frequency band (delta, theta, alpha, beta, gamma) to derive frequency-specific spatial filters. For each frequency band, let $E_n \in \mathbb{R}^{C \times T}$ represent the EEG data for trial $n$, where $C$ is the number of channels and $T$ is the number of time samples. For each class, the average trial covariance matrices for all trials in class 1 ($C_1$) and class 2 ($C_2$) are computed as $\Sigma^1$ and $\Sigma^2$ respectively, where $N_1$ and $N_2$ denote the number of trials within class 1 and class 2. These covariance matrices are calculated as follows:

\vspace{-7pt}
\begin{equation}
\Sigma^{1} = \left( \frac{1}{N_{1}} \right) \sum_{n \in C_{1}} \frac{(E_n E_n^{\mathrm{T}})}{\mathrm{trace}(E_n E_n^{\mathrm{T}})}
\end{equation}
\vspace{-7pt}
\begin{equation}
\Sigma^{2} = \left( \frac{1}{N_{2}} \right) \sum_{n \in C_{2}} \frac{(E_n E_n^{\mathrm{T}})}{\mathrm{trace}(E_n E_n^{\mathrm{T}})}
\end{equation}

\vspace{-3pt}
The eigenvector ($U$) and eigenvalues ($D$) obtained from the eigenvalue decomposition of the composite covariance matrix $\Sigma^{c}$ are used to compute the whitening transformation matrix $P$:
\vspace{-6pt}

\begin{equation}
\Sigma^{c} = \Sigma^{1} + \Sigma^{2} = U \, D \, U^{\mathrm{T}}, \quad P = D^{-\frac{1}{2}} \times U^{\mathrm{T}}
\end{equation}

\vspace{-5pt}
Using the whitening transformation matrix $P$, the average covariance matrices for both classes are transformed to have unit variance and zero covariance:
\vspace{-3pt}

\begin{equation}
\hat{\Sigma}^{1} = P \, \Sigma^{1} P^{\mathrm{T}}, \quad \hat{\Sigma}^{2} = P \, \Sigma^{2} P^{\mathrm{T}}
\end{equation}

\vspace{-5pt}
The shared eigenvector $V$ between the transformed covariance matrices $\Sigma^{1}$ and $\Sigma^{2}$ is used to compute the spatial filter matrix $W$ ($16 \times 16$) below. The preprocessed single-trial EEG data $S_n$ for both classes were projected onto all 16 spatial filters to obtain the filtered trial:
\vspace{-3pt}

\begin{equation}
W = P^{\mathrm{T}} V, \quad S_{n} = W^{\mathrm{T}} E_{n}
\end{equation}

\vspace{-5pt}

For each projected trial, the variance for each EEG channel is computed. Channel variances are normalized by the total variance summed across all channels and transformed using the logarithmic scale. These values are arranged to form the following feature vector for each trial, represented as $F_{\text{trial}} = [x_1, x_2, \ldots, x_{16}]$. Feature sets resulted in different dimensionalities depending on the analysis approach:
\begin{itemize}
    \item{Frequency-based:} 16 features per band ($1 \times 16$ vector)
    \item{Broadband:} Concatenated features across all bands ($1 \times 80$ vector)
\end{itemize}

\begin{figure*}[!t]
\centering
\includegraphics[width=0.9\textwidth]{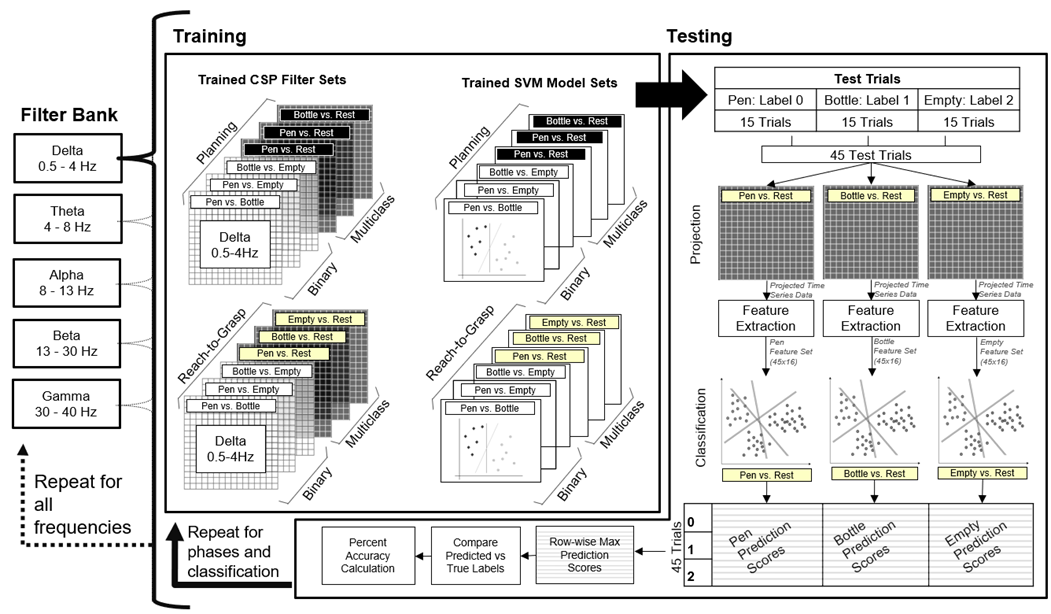}
\caption{Overview of FBCSP multiclass classification pipeline using one-vs-rest strategy with three trained binary classifiers for final class prediction.}
\label{fig:pipeline}
\end{figure*}

\subsubsection{FBCSP Filter Sets}
In the binary classification paradigm, CSP filters were computed using training trials to distinguish between three pairwise scenarios:

\noindent\textbf{Precision vs. power grasp scenario}
\begin{itemize}
   \item Pen (60 trials) vs. bottle (60 trials)
\end{itemize}
\textbf{Grasp vs. no grasp scenarios}
\begin{itemize}
   \item Pen (60 trials) vs. empty (60 trials)
   \item Bottle (60 trials) vs. empty (60 trials)
\end{itemize}

For multiclass classification, CSP filters were extended to handle all three conditions simultaneously using the one-vs-rest strategy~\cite{ang2008filter} given the binary constraint of the FBCSP method. CSP filters were computed using training trials to distinguish between three one-vs-rest scenarios:
\begin{itemize}
    \item Pen (60 trials) vs. rest (120 trials: bottle/empty)
    \item Bottle (60 trials) vs. rest (120 trials: pen/empty)
    \item Empty (60 trials) vs. rest (120 trials: pen/bottle)
\end{itemize}
FBCSP filter sets were generated for each classification scenario across all frequency bands (delta, theta, alpha, beta, gamma) and both task phases (planning, reach-to-grasp), resulting in $3 \text{ (scenarios)} \times 5 \text{ (frequency bands)} \times 2 \text{ (phases)} = 30$ individual binary CSP filters (Fig.~\ref{fig:pipeline}).

\vspace{-5pt}
\subsection{Movement-Related Cortical Potential (MRCP)} 
To extract MRCP features, a low-pass filter with a cutoff frequency of 6~Hz was applied to the preprocessed training EEG signals, isolating the slow cortical potentials associated with MRCPs. The grand average across all training trials was computed to derive the corresponding MRCPs for each class (pen, bottle, and empty). Temporal amplitude features of the MRCP were extracted from a 1.5-second window of interest (WOI), spanning 600~ms before to 1000~ms after reach-to-grasp onset. This window was divided into seven 200-ms intervals (7 total windows), in which the mean amplitude for each EEG channel was calculated for each interval. This resulted in a normalized $16 \times 7$ feature matrix for each class.

\vspace{-6pt}
\subsection{Classification}
\subsubsection{Training FBCSP-SVM Classifiers}

Individual binary SVM classifiers are trained separately based on the classification scenario (precision vs. power grasp, grasp vs. no grasp), frequency band, and phase, using the corresponding trained feature sets. Each SVM classifier was trained using 10-fold cross validation (CV), where feature standardization was applied to each fold on the training set to prevent data leakage. Model parameters were selected based on the maximum accuracy score across the folds. 

For multiclass classification, the one-vs-rest strategy was used. The classes are represented as $C = \{c_1, c_2, \ldots, c_K\}$, where $K = 3$. For each class $c_K$, a binary classifier $f_k(x)$ is trained to distinguish the target class (e.g., pen) from the combined set of the other two classes (e.g., bottle and empty). This resulted in three independent classifiers trained on the corresponding feature set (Fig.~\ref{fig:pipeline}). 

All finalized binary and multiclass classification models with corresponding weights and trained scalers were saved for feature importance analysis and evaluation on the unseen test dataset.

\subsubsection{Testing FBCSP-SVM Classifiers}

For multiclass tasks, 45 test samples (15 trials per class) were projected through the trained CSP filters to produce feature matrices of size $45 \times 16$. The one-vs-rest SVM classifiers generated confidence scores compiled into a prediction score matrix $S(x) = [s_1(x), s_2(x), \ldots, s_K(x)]$. Final classification was determined by selecting the class $k$ with the highest confidence score $s_k(x)$, where $k \in \{1, 2, \ldots, K\}$. Performance was evaluated by comparing predicted labels with ground truth labels.

\subsubsection{Training and Testing of MRCP-SVM Classifiers}

Individual MRCP classifiers were trained using 10-fold CV with class-specific feature sets of 16 channels $\times$ 7 features. Binary classification used stacked feature sets ($32 \times 7$ matrix), while multiclass classification combined all three classes ($48 \times 7$ matrix). 

\subsubsection{Performance Evaluation and Statistical Analysis}

Classification performance was evaluated using 10-fold CV on the training set for binary classification tasks and on the unseen test set for multiclass classification using a one-vs-rest approach. Both frequency-specific and broadband feature sets were assessed across all classification scenarios and task phases, with unseen test results reported as mean $\pm$ standard deviation across subjects. The significance threshold is set at $p < 0.05$ for comparisons between task phases and frequency bands using paired t-tests across all subjects. Feature importance was evaluated by extracting SVM coefficients from all finalized models to identify the most discriminative spatial patterns. FBCSP spatial filters and their corresponding projected data were saved and visualized through topographical maps to examine the emergence of discriminative neural patterns across different experimental conditions.

\begin{figure*}
\centering
\includegraphics[width=0.7\textwidth]{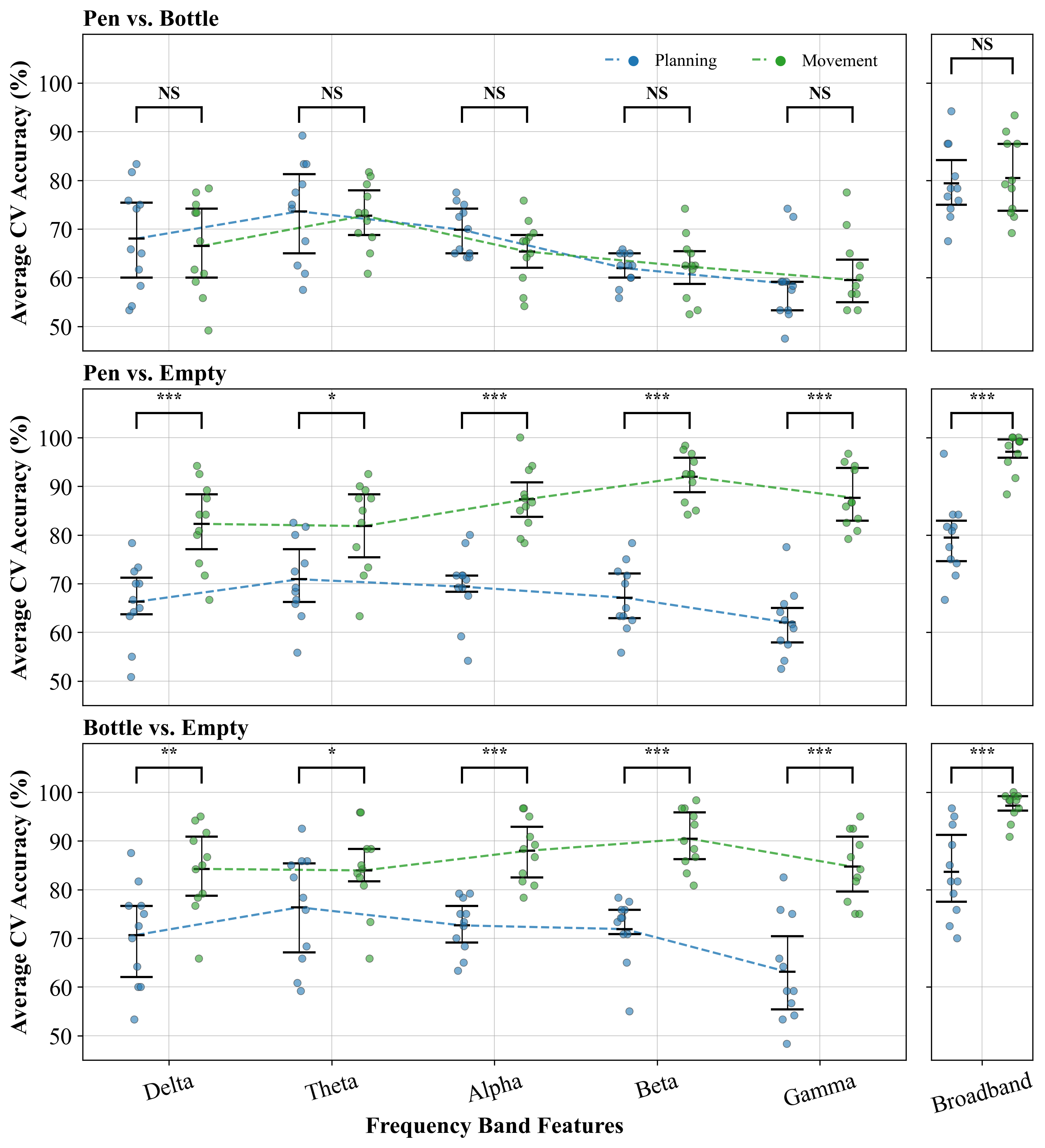}
\caption{Classification accuracies for cross-validation analysis across different experimental conditions and EEG frequency bands. Results show average classification performance ($\pm$STD) across all subjects using 10-fold CV on the training set for three binary conditions: pen vs. bottle (top), pen vs. empty (middle), and bottle vs. empty (bottom). Blue circles represent planning phase data, green circles represent movement phase data, with individual frequency bands (delta, theta, alpha, beta, gamma) and broadband features (concatenated across all bands) shown on the x-axis. Statistical significance between task phases are indicated by asterisks (*$p < 0.05$, ***$p < 0.001$; NS = not significant).}
\label{fig:classification}
\end{figure*}

\section{Results}
\subsection{Classification Results}

\subsubsection{Frequency-Based Binary and Multiclass Classification Performance}

Figure~\ref{fig:classification} shows FBCSP classification performance ($\pm$STD) across all subjects using the average across 10-fold CV for three binary scenarios: pen vs. bottle (top), pen vs. empty (middle), and bottle vs. empty (bottom). Each panel plots average CV Accuracy (50-100\%) against frequency band features (delta, theta, alpha, beta, gamma, broadband). 

During the planning phase, theta band features showed the highest classification performance across all binary scenarios. Average performance across subjects reached 75.3\% for pen vs. bottle, 70.4\% for pen vs. empty, and 77.6\% for bottle vs. empty scenario. Following peak theta performance, there is a downward trend in classification performance across the frequency spectrum for all binary scenarios. As an example, pen vs. bottle performance peaked in lower frequencies-delta (69.3\%) and theta (76.2\%)-then declined through alpha (69.1\%), beta (62.3\%), and gamma (57.7\%) (theta $>$ beta, gamma; alpha $>$ beta, gamma; $p < 0.05$). For the pen vs. bottle movement execution phase, theta achieved 73.9\% accuracy, significantly higher than alpha, beta, and gamma bands ($p < 0.05$). However, in the grasp vs no grasp movement phase scenarios, beta features reached peak performance at 92.3\% (pen vs. empty) and 90.4\% (bottle vs. empty). Across the frequency spectrum, there is an upward trend in classification performance in the higher frequency bands following the delta and theta bands, opposite to the downward trend seen during planning. 

The relationship between planning and movement phases depended on both frequency band and scenario. For the pen vs. bottle scenario, statistical analysis showed no significant differences between phases across any frequency band ($p > 0.05$). However, grasp vs. no grasp scenarios showed significant increases between grasp planning to movement execution in higher frequencies. Pen vs. empty accuracy increased in alpha (87.6\%), beta (93.3\%), and gamma (87.5\%) ($p < 0.05$), while bottle vs. empty showed similar increases in alpha (87.5\%), beta (89.6\%), and gamma (84.4\%) ($p < 0.05$).

The FBCSP spatial filters used for feature extraction and classification in Figure~\ref{fig:classification} derived from the training set were evaluated on an unseen test set to assess their ability to capture discriminative features. Table~\ref{tab:classification_results_multiclass} reports the average ($\pm$ STD) multiclass classification performance for all scenarios, frequency bands, and phases. Similarly, all binary scenarios achieved peak theta performance during grasp planning ($65.2 \pm 11.2\%$ (pen vs. bottle), $58.2 \pm 13.1\%$ (pen vs. empty), and $67.3 \pm 17.4\%$ (bottle vs. empty). The downward trend in classification performance across the frequency spectrum can also be seen as accuracy decreases to $43.9 \pm 9.3\%$, $50.6 \pm 7.6\%$, and $52.1 \pm 8.3\%$ in the gamma band, respectively. Peak beta performance occurs in grasp vs. no grasp movement phase scenarios, with $88.2 \pm 7.7\%$ (pen vs. empty) and $86.7 \pm 6.7\%$ (bottle vs. empty), whereas pen vs. bottle movement phase achieved below chance level at  $47.0 \pm 8.8\%$. 

\subsubsection{Broadband Binary and Multiclass Classification Performance}

When using all frequency features simultaneously (80 features), broadband classification achieved the highest performance across both grasp planning and execution phases in all binary scenarios when compared to only using frequency band-specific features (16 features) (Fig.~\ref{fig:classification}). Pen vs. bottle showed no significance between grasp planning (79.7\%) and movement execution (81.5\%) phase accuracy, while grasp vs. no grasp scenarios demonstrated significant phase differences with movement accuracies reaching 96.4\% (pen vs. empty) and 97.2\% (bottle vs. empty) compared to planning accuracies of 78.6\% and 83.7\%, respectively ($p < 0.05$). MRCP-based classification achieved lower performance $61.1 \pm 7.2\%$ (pen vs. bottle), $68.5 \pm 5.4\%$ (pen vs. empty), and $66.3 \pm 6.5\%$ (bottle vs. empty). Table~\ref{tab:classification_results_multiclass} compares multiclass FBCSP grasp planning/movement execution results using broadband features to the established MRCP technique on the unseen test set. Based on Table 1, using FBCSP broadband features led to improved performance when compared to MRCP across all scenarios and phases. For multiclass classification, FBCSP broadband achieved 67.4±10.6\% (grasp planning) and 78.3±4.4\% (movement execution) versus MRCP performance at 52.7±8.7\%.

\begin{figure*}[!t]
\centering
\includegraphics[width=\textwidth,trim=0 0 0 85,clip]{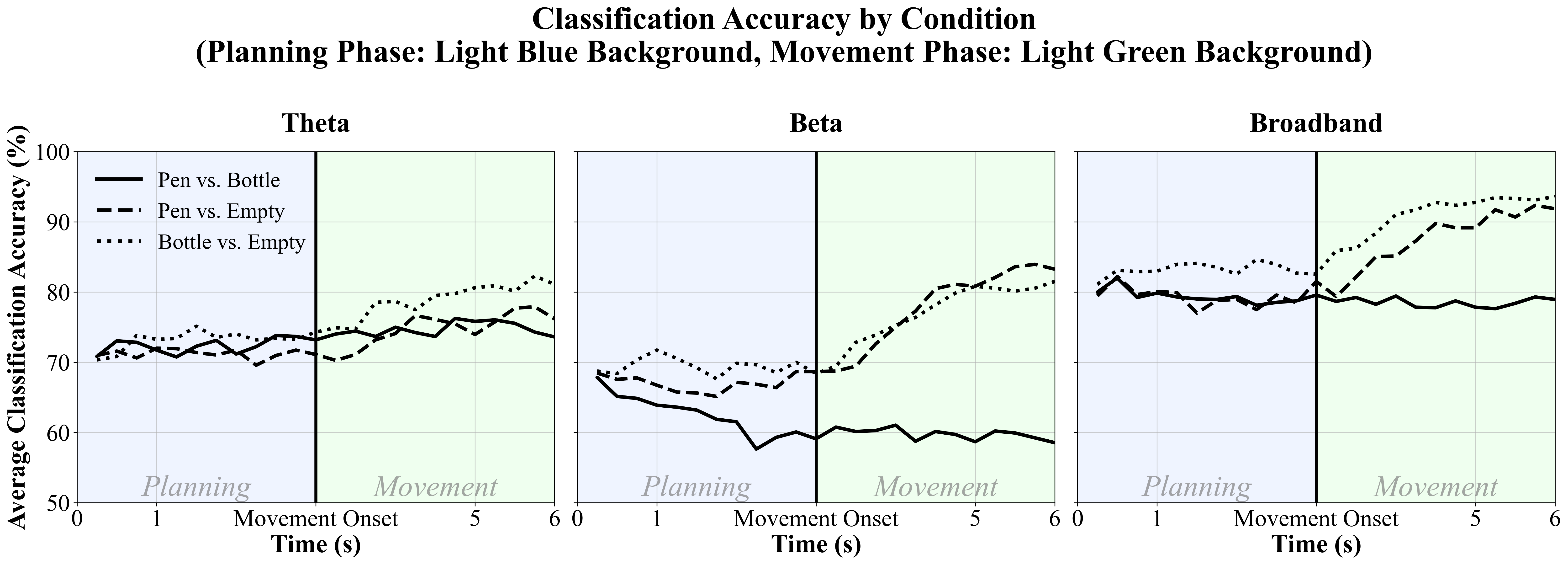}
\caption{Temporal evolution of 10-fold CV classification accuracy across frequency bands during plan-to-grasp tasks. Classification performance is shown over the plan-to-grasp task within three frequency bands (theta, beta, Broadband) and three binary conditions (pen vs. bottle, pen vs. empty, bottle vs. empty), averaged across all subjects. The number of extracted features remains constant at each time point, with classification performance varying based on the cumulative temporal information available for feature extraction. The vertical line at movement onset (t = 3s) separates the planning phase (blue shaded region) from the movement phase (green shaded region).}
\label{fig:temporal_classification}
\end{figure*}

\subsubsection{Temporal Evolution of Selected Frequency Bands and Broadband Classification}

To assess temporal emergence of classification accuracy, performance was analyzed across three frequency bands (theta, beta, broadband) for three binary conditions (pen vs. bottle, pen vs. empty, bottle vs. empty), averaged across all subjects (Fig.~\ref{fig:temporal_classification}). Classification windows incrementally expanded from time zero, extracting new features at each time point to evaluate how accuracy evolved with increasing temporal information. The classification accuracy using theta band features remained relatively stable within the  72-76\% range across all binary scenarios from the grasp planning phase through movement execution phases. Statistical analysis revealed no significant phase differences for pen vs. bottle and pen vs. empty conditions, while bottle vs. empty showed significant differences between planning and movement phases ($p < 0.05$). Beta band features showed similar planning accuracy trends at lower levels (~65-70\%) across all binary scenarios, but grasp vs. no grasp movement phase scenarios begin to trend upward to 80-85\% after movement onset, while pen vs. bottle remained at ~60\%, with significant planning-to-movement differences in empty conditions ($p < 0.05$) but not in pen vs. bottle ($p > 0.05$). Broadband features demonstrated similar patterns, with grasp vs. no grasp movement phase scenarios reaching 90-93\% peak performance while pen vs. bottle showed minimal change.

\vspace{-7pt}
\subsection{Emergence of Plan-to-Grasp Feature Discriminability}
\subsubsection{CSP Topographical Analysis}
The observed classification differences across scenarios, frequency bands, and phases reflect the discriminative neural features extracted by the FBCSP algorithm. To identify which features drove these patterns, feature importance was examined across classifiers to examine the spatial and spectral characteristics most informative for each decision. For representative Subject 8, CSP topographies and SVM coefficient maps illustrate the spatial distribution of discriminative features across conditions and phases (Fig.\ref{fig:subject8_topo}). Since peak accuracies emerged in the theta band during grasp planning and the beta band for grasp vs no-grasp movement phase scenarios (Fig.\ref{fig:classification}), the displayed CSP maps correspond to features with the highest SVM coefficients within those bands.

For the grasp planning phase, CSP patterns were with the maximum SVM coefficients within the theta band were displayed. The pen vs. bottle condition (CSP4) showed discriminative activation mainly across the left parieto-occipital region (PO7, P4, Oz) and frontal electrodes (FP1), with emerging activations from left motor cortex (C3). For the pen vs. empty (CSP2) and bottle vs. empty (CSP2) conditions, discriminative activation was concentrated in parietal (P3, Pz) and occipital regions.

For the grasp movement execution phase, CSP patterns within the beta band were selected. The pen vs. bottle condition (CSP9) showed strong discriminative activity over the left motor cortex (C3), with additional involvement of occipital and parietal regions (PO7, Oz, P4, Cz, F3, F4). Both pen vs. empty (CSP9) and bottle vs. empty (CSP1) showed activity at Cz and Fz, with pen vs. empty also engaging motor (C3) and parietal (P3, Pz) regions, while bottle vs. empty showed stronger occipital activity (PO8, Oz).

\begin{table}
\renewcommand{\arraystretch}{1.5}
\caption{Comparison of average multiclass classification accuracy (mean $\pm$ std) across subjects using one-vs-rest FBCSP with learned spatial filters (Planning / Movement) and the MRCP method (WOI) on held-out test data}
\label{tab:classification_results_multiclass}
\centering
\setlength{\tabcolsep}{16pt}
\begin{tabular}{|c|c|}
\hline
\textbf{Band} & \textbf{FBCSP (Planning / Movement)} \\
\hline
$\delta$ & 53.5$\pm$6.6 / 66.3$\pm$9.6 \\
\hline
$\theta$ & 56.6$\pm$11.9 / 69.3$\pm$7.0 \\
\hline
$\alpha$ & 52.8$\pm$8.3 / 67.6$\pm$8.3 \\
\hline
$\beta$ & 48.5$\pm$7.1 / 67.4$\pm$7.6 \\
\hline
$\gamma$ & 43.2$\pm$6.3 / 61.5$\pm$9.3 \\
\hline
Broadband & 67.4$\pm$10.6 / 78.3$\pm$4.4 \\
\hline
\multicolumn{2}{|c|}{\textbf{MRCP (Includes both planning and movement phases)}} \\
\hline
\textbf{WOI} & 52.7$\pm$8.7 \\
\hline
\end{tabular}
\end{table}

\subsubsection{SVM Coefficient Analysis and Feature Importance} 

\begin{figure*}
\centering
\includegraphics[width=\textwidth,trim=0 0 0 50,clip]{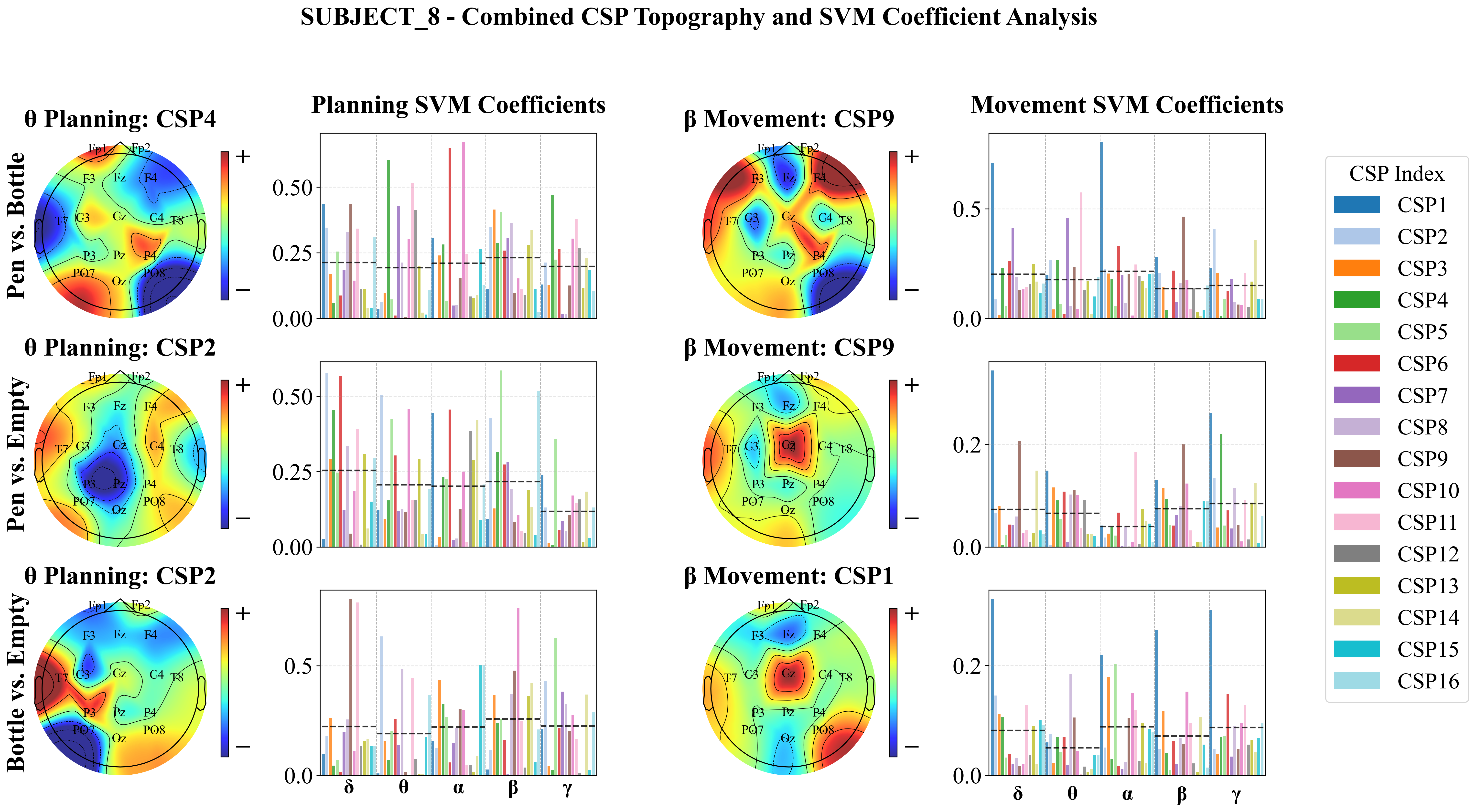}
\caption{Most significant CSP patterns and corresponding broadband SVM coefficients for representative Subject 8. The left panel shows the most significant CSP patterns in the theta band (CSP4 and CSP2) and their corresponding broadband SVM coefficients during the planning phase, while the right panel displays the most significant CSP patterns in the beta band (CSP9 and CSP1) and their corresponding broadband SVM coefficients during the movement phase. Topographic maps illustrate the spatial distribution of CSP patterns with positive (red) and negative (blue) weights across electrode locations. Bar plots show the magnitude of broadband SVM coefficients across different CSP indices (CSP1-CSP16), with colored bars representing individual CSP components and dashed horizontal lines indicating mean values across each frequency band.}
\label{fig:subject8_topo}
\end{figure*}

\begin{figure*}
\centering
\includegraphics[width=0.85\textwidth,trim=0 0 0 30,clip]{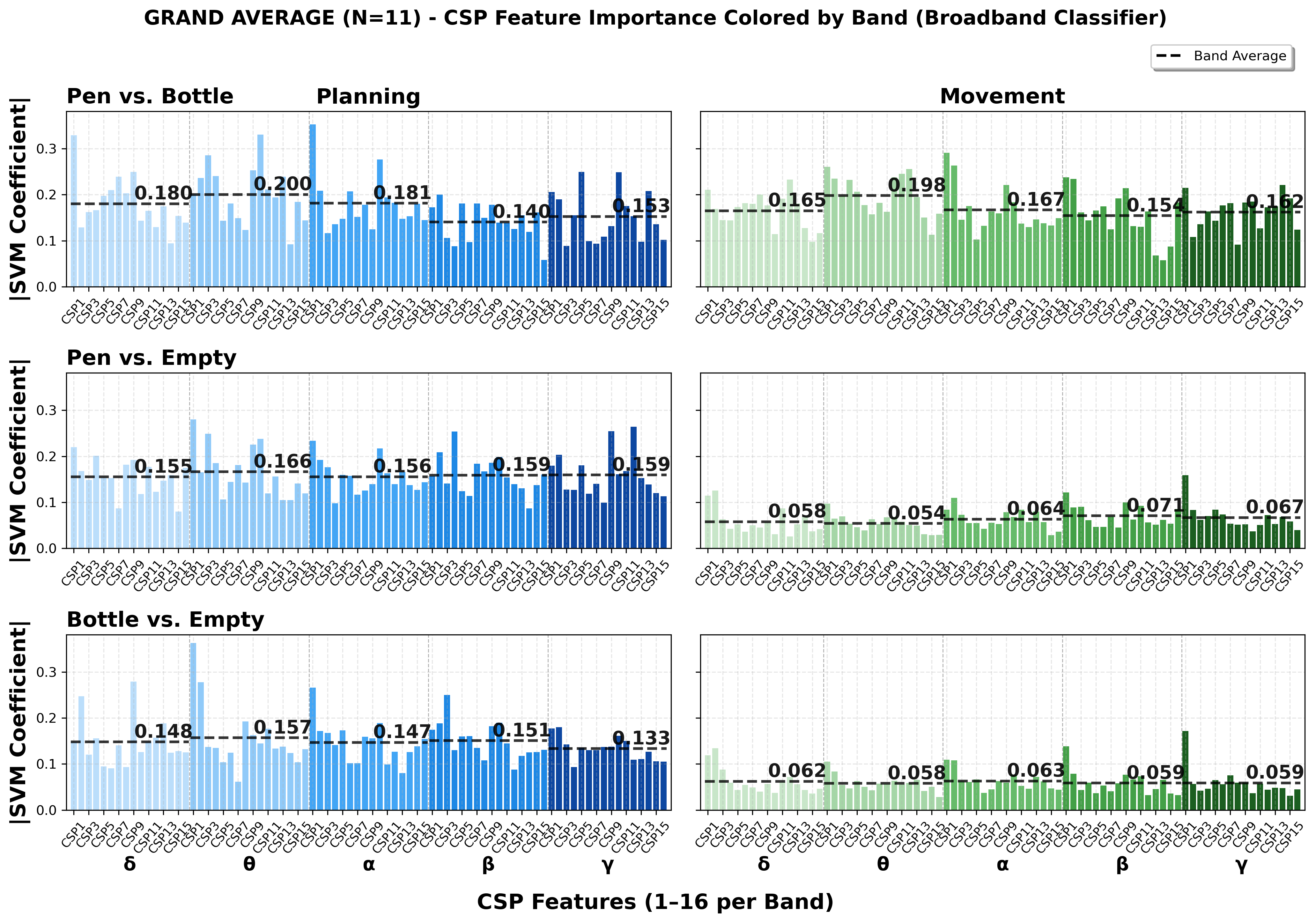}
\caption{Grand average absolute SVM coefficients for CSP features across all subjects during broadband classification. Left column shows planning period results, right column shows movement period results. Each bar represents one CSP feature (1-16 per frequency band) with height indicating feature importance. Dashed horizontal lines indicate the mean coefficient magnitude within each frequency band. Three classification comparisons are shown: pen vs. bottle (top row), pen vs. empty (middle row), and bottle vs. empty (bottom row).}
\label{fig:grand_average_csp}
\end{figure*}

Based on the mean frequency band coefficients in Fig.~\ref{fig:subject8_topo}, average SVM coefficients show an increase in value toward higher frequencies in grasp vs no grasp movement phase scenarios, indicating increased classifier reliance on high-frequency derived features. These results are similar to what was observed in the average classification results across subjects (Fig.~\ref{fig:classification}). Fig.~\ref{fig:grand_average_csp} presents a barplot of absolute SVM coefficient values averaged across subjects with respect to binary scenarios and phase. The analysis shows patterns analogous to Fig.~\ref{fig:classification} since theta band coefficients peaked at 0.200 (pen vs. bottle, theta vs. gamma: $p < 0.05$), 0.166 (pen vs. empty), and 0.157 (bottle vs. empty) during grasp planning across all scenarios alongside the downward trend of SVM coefficient values following the theta peak as frequencies increase. Conversely, grasp vs no grasp movement phase scenarios demonstrate an upward trend toward higher frequencies.

\vspace{-10pt}
\subsection{Temporal Dynamics Across Projected and Filtered Domains}

\subsubsection{Temporal Evolution using FBCSP}
\begin{figure}
\centering
\includegraphics[width=\columnwidth]{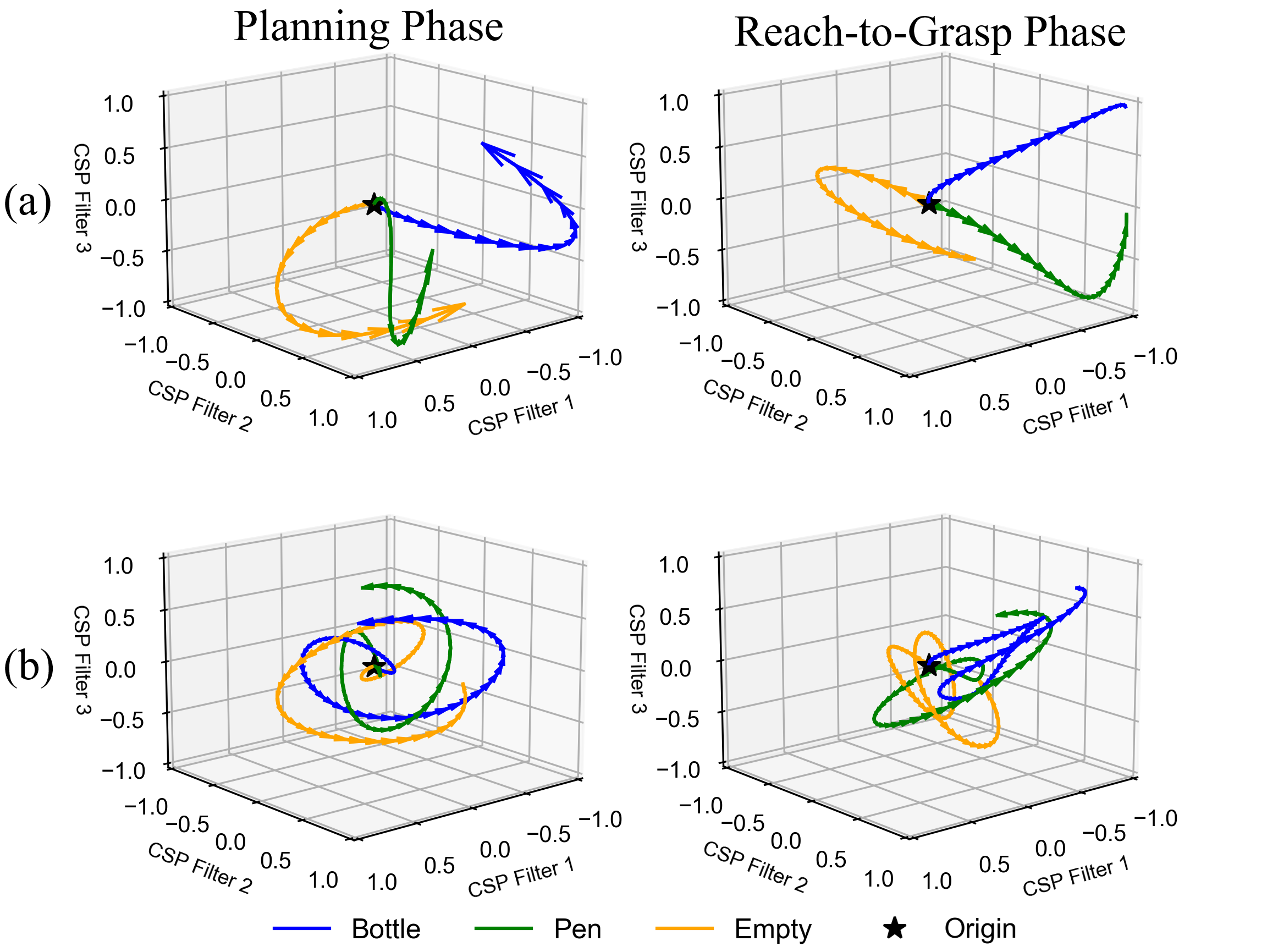}
\caption{Temporal evolution of neural trajectories in CSP filter space (theta band) for representative participant 8 in the multiclass framework. Left: planning phase, Right: reach-to-grasp phase at (a) 100ms and (b) 400ms into each phase.}
\label{fig:csp_trajectory_subj3_theta}
\end{figure}

Fig.~\ref{fig:csp_trajectory_subj3_theta} visualizes the normalized reconstructed trajectories of grand-averaged training trials for a representative participant (subject 8) projected into the CSP filter space (theta band) to display the neural dynamics during the plan-to-grasp task. The left and right columns represent grasp planning and movement execution phases, respectively, with the projected time series data using CSP components \#1, \#2, and \#3 mapped to the x, y, and z axes. These CSP filters were trained in a multiclass framework (one vs. rest approach) to distinguish between one class versus the rest (bottle vs. rest, pen vs. rest, empty vs. rest), maximizing the variance of one class while minimizing the variance of others to create a 3D visualization of spatial separation between individual grasping intentions. The rows in Fig.~\ref{fig:csp_trajectory_subj3_theta} show the temporal progression within each phase at $\sim$100~ms (a) and $\sim$400~ms (b). All paths were scaled to originate from (0,0,0), marked by a black star. During the planning and reach-to-grasp phase, the trajectories display initial divergence across all conditions at 100~ms. During the planning phase at 400~ms the trajectories form clear circular and looping patterns and begin to converge, whereas during the movement execution phase, the circular trajectories begin to overlap as the grasping action unfolds.

\subsubsection{Temporal Evolution using MRCP}
\begin{figure}
\centering
\includegraphics[width=\columnwidth]{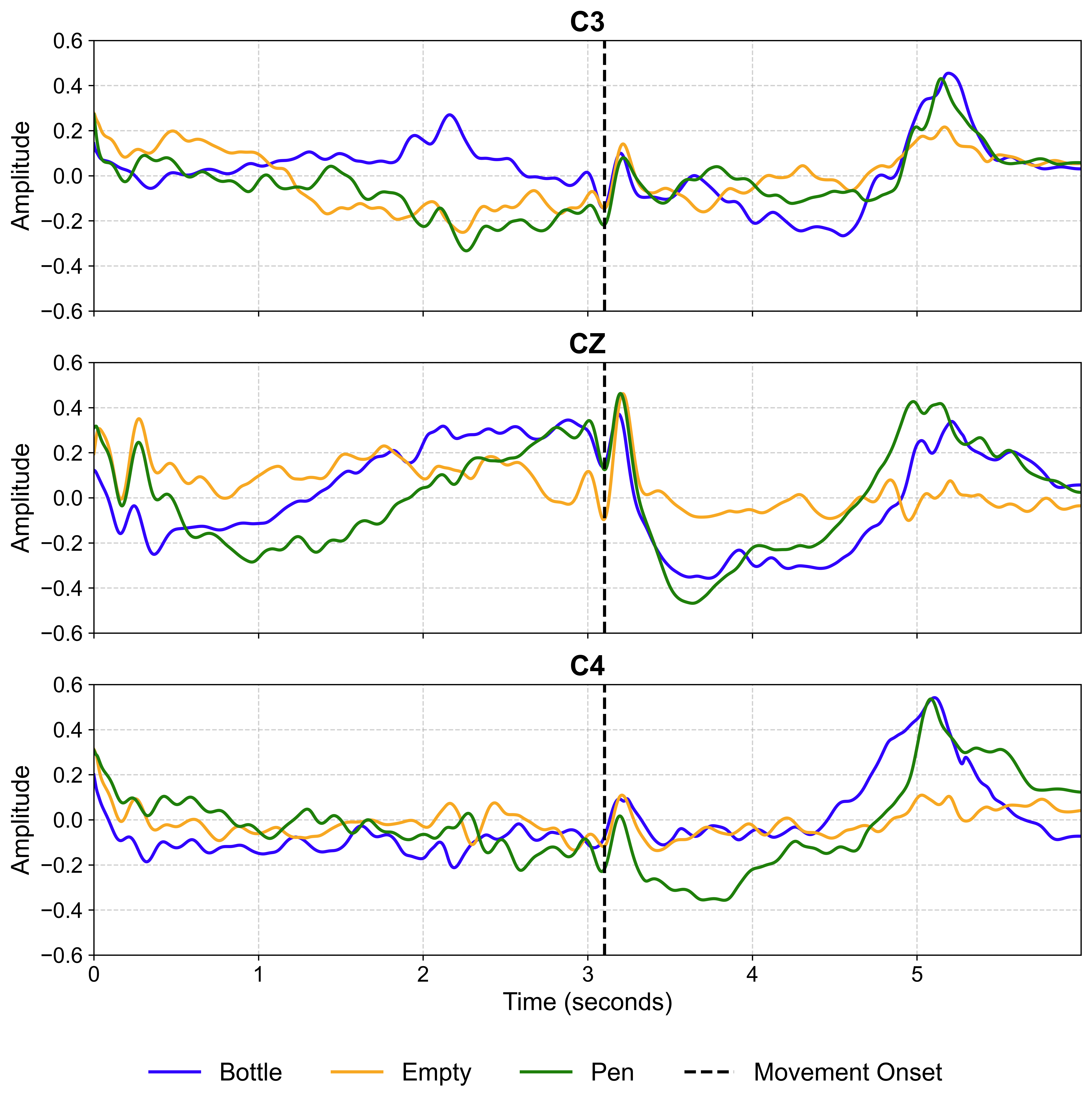}
\caption{Normalized grand average MRCP waveforms for representative participant 8 at C3, Cz, and C4. The dashed line marks reach-to-grasp onset, while with colored potentials represent pen (blue), bottle (purple), and empty (green) conditions.}
\label{fig:mrcp}
\end{figure}

Fig.~\ref{fig:mrcp} shows the MRCP waveforms for one participant across three motor-related electrodes (C3, Cz, and C4) where each color represents the individual object scenario (pen, bottle, and empty). The x-axis represents time in seconds and the y-axis measures normalized amplitude while movement onset is marked by a vertical dashed line. Across all three electrode sites, the characteristic Bereitschaftspotential (BP) is shown as a negative deflection occurring just before movement onset. The BP at C3 and Cz has a negative shift followed by a positive rebound after movment onset. Similar, albeit less pronounced, negative deflections are visible at C4. The pen condition showed the strongest BP at Cz, while bottle and empty conditions displayed smaller negative deflections. All conditions exhibited positive rebound post-execution, consistent with reach-to-grasp cortical potentials.

\vspace{1\baselineskip}    
\section{Discussion}

\vspace{-1pt}
\subsection{Role of Low-Frequency Oscillations in Grasp Planning}

By extracting features and classifying the isolated planning and movement phases of plan-to-grasp tasks using FBCSP, different layers of a spectral profile of the complex vision-based grasping network underlying visuomotor tasks were captured. The CSP topographical analysis (Fig. ~\ref{fig:subject8_topo}) show discriminative theta patterns in parieto-occipital and frontal regions during planning, similar to the findings Riddle et al.~\cite{riddle2020oscillatory} and Pagnotta et al.~\cite{pagnotta2024multiplexed}, who demonstrated that theta oscillations (4-7 Hz) in frontal and parietal networks reflect the maintenance of rules linking stimuli to appropriate actions. The features extracted from theta oscillations peak in classification accuracy for both planning (75.3\%) and movement (73.9\%) phases for precision vs. power grasp discrimination ~\ref{fig:classification}. In the same grasp vs. grasp scenario, the classification accuracy, using features extracted from alpha, beta, and gamma oscillations decreased for both planning and movement phases when compared to theta performance. Similarly, when looking at the SVM coefficient analysis (Fig.~\ref{fig:grand_average_csp}), CSP features extracted from theta oscillations had the greatest coefficient magnitudes across all scenarios during planning compared to other frequency bands. During the movement phase of the precision vs. power grasp scenario, CSP features extracted from alpha, beta, and gamma were lower in magnitude. This may suggest that precision and power grasps rely on the same sensorimotor patterns that are embedded in higher frequencies, leaving theta-driven cognitive control as the key discriminator.

In contrast, when classifying either precision or power grasp with respect to no movement, there is a phase-dependent shift of frequency importance, where higher frequency oscillations become dominant in the transition from planning to movement execution (Fig. ~\ref{fig:classification}). While theta oscillations dominated during planning phases across all scenarios, classification accuracies peaked at 93.3\% (pen vs. empty) and 89.6\% (bottle vs. empty) during movement when using features extracted from beta oscillations, while gamma features reached 87.5\% and 84.4\%, respectively. The emergence of beta and gamma oscillations during grasp vs. no grasp movement execution scenarios likely originates from the integration of the sensorimotor network recruited for object handling, which is associated with visuomotor tasks found within the 12-35 Hz frequency range that includes the mu rhythm~\cite{pfurtscheller2003motor}. The CSP topographical analysis (Fig. ~\ref{fig:subject8_topo}) shows discriminative patterns above the central electrodes (specifically Cz) within the beta band, also associated with the location where the mu rhythm is found.

The importance of higher frequencies emerges when classifying either grasp type vs. no movement, capturing cognitive processes related to higher-order attention, task engagement, and sensorimotor integration associated with visuomotor tasks, rather than grasp-specific information due to the shared motor demands between precision and power grasps ~\cite{khanna2015_betamotor, schmidt2019beta}. However, power vs. precision grasp classification shows peak theta performance in both phases and decreased accuracy when using features within higher frequencies, suggesting that grasp-specific discriminative information resides primarily in low-frequency oscillations.

\vspace{-10pt}
\subsection{Evidence of Temporal Macroscopic Emergence in Plan-to-Grasp}

Our findings provide evidence for the temporal emergence of discriminative neural patterns that align with established visuomotor networks underlying vision-based grasping. The CSP topographical analysis (Fig.~\ref{fig:subject8_topo}) shows the transition of discriminative activity from posterior visual processing areas during planning to motor execution regions during movement, reflecting the known cortical hierarchy of grasp control networks~\cite{rizzolatti2014cortical, filimon2010human}. During the planning phase, discriminative theta patterns are predominantly localized to parieto-occipital regions (PO7, P4, Oz) and frontal electrodes (FP1), consistent with the dorsal visual stream's role in transforming visual object information into motor intent~\cite{culham2006human, cisek2007cortical}. The pen vs. bottle condition showed discriminative activation in the left parieto-occipital region with emerging patterns from the left motor cortex (C3), while object-versus-empty conditions demonstrated concentrated activity in parietal (P3, Pz) and occipital regions. This may relate to primate neurophysiology, where neurons in areas AIP and F5 exhibit grip-selective responses during object presentation~\cite{murata2000selectivity, baumann2009context}. In humans, the homologous anterior intraparietal sulcus (aIPS) and ventral premotor cortex (PMv) are similarly engaged during action observation, object perception, and delayed grasp planning. Corresponding macroscopic activity is captured by parietal (P3/P4) and central (C3) electrodes, reflecting the involvement of the aIPS and PMv networks. Combined TMS-EEG studies confirms the functional relevance of these sites, showing that disruption of aIPS significantly reduces muscle-specific PMv-M1 interactions during grasp preparation~\cite{davare2010causal}. 

Supporting this, classification accuracy in the theta band remained within the 70-76\% range from grasp planning through movement execution across all scenarios (Fig.~\ref{fig:temporal_classification}). This stability suggests that theta-band activity encodes grasp-relevant information during planning, and that adding movement-related temporal data does not increase classification performance; indicating no additional discriminative information is gained beyond the planning phase. In contrast, classification performance began to significantly increase ($p < 0.05$) when using features extracted from beta oscillations only after movement onset in the grasp vs no grasp scenarios. This may indicate that beta activity corresponds to execution-related neural processes that emerge during actual object manipulation~\cite{schmidt2019beta}. Broadband classification combines all features (80) across the entire frequency spectrum (0.5 - 40 Hz). In the precision vs. power grasp condition, including higher-frequency features resulted in only 3-4\% improvement in classification over theta alone (16 features), with performance remaining flat across the trial. However, for grasp vs. no-grasp scenarios, an increase in classification accuracy emerges only after movement onset, showing the importance of movement-specific high-frequency activity in those scenarios.

These findings demonstrate the temporal macroscopic emergence of distinct neural processes in plan-to-grasp tasks, where low-frequency theta activity carries grasp-specific information during the planning phase before movement begins, while beta oscillations emerge later to contribute a different temporal layer of the vision-based grasping network related to task engagement and sensorimotor processing rather than grasp-specific discrimination. This temporal separation shows how macroscopic neural patterns emerge sequentially, with planning-related grasp information encoded early in theta oscillations and execution-related processes in higher frequencies.

\vspace{-10pt}
\subsection{Towards Naturalistic BMI Control for Plan-to-Grasp Tasks}

Our results provide evidence that EEG-based grasp decoding during the planning phase could be a viable strategy for naturalistic BMI control in individuals with SCI. Unlike prior EEG studies which focused on reach-to-grasp decoding using MRCPs~\cite{iturrate2018_eeggrasp}, our novel platform is designed to isolate neural activity during grasp planning before any movement occurs using objects involved in ADL. Invasive studies show that grasp planning activates neural circuits involved in object recognition, geometric feature extraction, and grasp type selection, allowing the transformation of visual input into motor intentions without actual movement~\cite{rizzolatti2014cortical}. This temporal separation of cognitive processes associated with grasp planning and motor processes involved during grasp execution could be advantageous, as it may introduce a new naturalistic control strategy for BMI systems, as patients with severe motor impairments require systems that bypass impaired motor pathways and instead decode cognitive signals~\cite{chinellato2009_neuroscience, abiri2019comprehensive}.

A major limitation of prior studies in noninvasive grasp decoding is their reliance on paradigms and analysis techniques that combine neural information from both planning and movement phases. This prevents the exploration of an isolated, yet naturalistic and intuitive way to explore grasp planning neural dynamics for movement-free neuroprosthetic control~\cite{iturrate2018_eeggrasp, schwarz2020_analyzing}. A common method in these paradigms, the MRCP approach (while successful for distinguishing grasp types~\cite{shiman2017_classification}) captures neural activity from both the planning and movement execution periods within its temporal window of interest (-2 to +1 seconds relative to movement onset). Since MRCP is limited to slow cortical potentials ($<$ 6 Hz), it fails to capture the full spectral information across frequency bands that may provide a comprehensive understanding of how the macroscopic vision-based grasping system networks emerge during different phases for specific grasp types. For example, Schwarz et al.~\cite{schwarz2020_analyzing} and Sburlea et al.~\cite{sburlea2021_disentangling} used MRCPs for classifying grasp types in reach-to-grasp tasks, achieving 62.3\% (lateral vs. palmar vs. rest) and 33.0-52.8\% accuracy (power vs. precision (two finger) vs. precision (five finger)), respectively. While successful in reach-to-grasp decoding, these paradigms used simultaneous object presentations, limiting the ability to map neural activity to specific object-grasp pairings during grasp planning. Similarly, Xu et al.~\cite{xu2021_decoding} and Iturrate et al.~\cite{iturrate2018_eeggrasp} achieved similar decoding performance using MRCP features, but required executed movements, leaving the role of planning unexplored. 

In contrast, our FBCSP-based method segments planning from execution and shows that theta-band features alone during planning reached 75.3\% classification accuracy (precision vs. power grasp) compared to MRCP performance (61.1\%) for the same task. Adding broadband features (80 total) only slightly improves accuracy (+4.4\%) during precision vs. power grasp planning classification, suggesting theta-band decoding offers a computationally efficient solution. This aligns with our previous findings that low-frequency oscillations encode grasp type for neuroprosthetic control~\cite{silversmith2021plug}, yet our noninvasive approach captures this during the planning phase, without surgical intervention. Reduced feature dimensionality further lowers overfitting risks and improves generalization (as seen in our unseen test set for multiclass classification), which is commonly a challenge when translating laboratory BMI systems to clinical applications.

\vspace{-8pt}
\subsection{Limitations and Future Work}

While this study demonstrated the feasibility of decoding grasp planning types from EEG, several limitations should be considered. Although the analysis successfully isolated and classified grasp planning signals (3-second window), it did not include detailed kinematic tracking of the reach-to-grasp movements to capture the full complexity of this multi-phase motor execution. The current plan-to-grasp task can further be broken down into smaller temporal windows—initially driven by visual and proprioceptive planning mechanisms, followed by real-time sensorimotor integration for trajectory control, hand shaping, and force modulation. These processes are supported by coordinated activity across parietal, premotor, and sensorimotor regions~\cite{filimon2010human, culham2006human} and may manifest in different frequency bands over time, as shown in this study. Therefore, future work should expand electrode coverage to explore a more detailed mapping of how motor commands are planned, refined, and executed across time. 

Combining high-resolution simultaneous neural and kinematic and behavioral data, such multimodal approaches and data availability may introduce new and more advanced algorithms, such as AI-inspired machine learning models to better decode brain network function. Incorporating neuro-inspired AI approaches, particularly those modeling ventral and dorsal stream processing involved in the vision-based grasping network, will advance our understanding of intention-to-action transformations. These advanced paradigms may soon introduce the possibility of real-time, intention-driven BMI control strategies that are more adaptive, intuitive, and functionally relevant for SCI patients.

\section{Conclusion}

In this study, a novel EEG-based platform was developed that temporally isolates grasp planning from execution phases during naturalistic object interactions, allowing phase-specific analysis of vision-based grasping neural dynamics. Through macroscopic-level analysis using FBCSP to classify discriminative frequency-specific neural activity, it was demonstrated that theta oscillations (4-8 Hz) contain grasp-specific information during planning phases that are maintained into the grasp execution phase when classifying between precision vs. power grasp discrimination. Whereas higher frequency bands (beta, gamma) showed decreased performance for grasp-type classification but became dominant during object manipulation phases when classifying between either power or precision grasp vs. no grasp, showing another layer of the complex vision-based grasping through sensorimotor integration rather than grasp-specific control. Our new FBCSP approach improved classification performance compared to established MRCP-based methods since FBCSP identified frequency-specific discriminative features within each phase, showing activity across parieto-occipital and frontal networks during planning that transitions to motor cortical areas during execution. These findings demonstrate that grasp intentions can be decoded most accurately from low-frequency oscillations prior to movement onset, establishing a foundation for naturalistic vision-based grasping control strategies for BMI applications.

\section*{Acknowledgment}

This research study was supported by the NSF CAREER under award ID 2441496 and the NSF grant under award ID 2245558. Additionally, the project was supported by URI Foundation Grant on Medical Research and the Rhode Island INBRE program from the National Institute of General Medical Sciences of the NIH under grant number P20GM103430.

\bibliography{refs}

\end{document}